\documentclass[review, 3p, sort&compress]{elsarticle}

\usepackage{lineno}
\usepackage{amssymb}
\usepackage{xspace}
\usepackage{xcolor}
\journal{Surface and Coatings Technology}

\newcommand{\singleColumnWidth}{8cm}
\newcommand{\uu}[1]{\ensuremath{\,\mathrm{#1}}}
\newcommand{\Bk}{B\textsubscript{k}\xspace}
\newcommand{\Bh}{B\textsubscript{h}\xspace}
\newcommand{\Bi}{B\textsubscript{i}\xspace}
\newcommand{\eps}{\ensuremath{\varepsilon}\xspace}
\newcommand{\corr}[1]{{\color{black}\protect#1}}

\begin{document}

\begin{frontmatter}
\title{Vacancy-driven extended stability of cubic metastable Ta-Al-N and Nb-Al-N phases}
\author[cdl]{Ferdinand Pacher}
\author[cdl,tuw]{Paul H. Mayrhofer}
\author[mul]{David Holec\corref{corr}}
\ead{david.holec@unileoben.ac.at}
\cortext[corr]{Corresponding author}
\address[cdl]{Christian Doppler Laboratory for Application Oriented Coating Development at the Institute of Materials Science and Technology, TU Wien, Getreidemarkt 9, A-1060 Vienna, Austria}
\address[tuw]{Institute of Materials Science and Technology, TU Wien, Getreidemarkt 9, A-1060 Vienna, Austria}
\address[mul]{Department of Physical Metallurgy and Materials Testing, Montanuniversi\"at Leoben, Franz-Josef-Stra\ss{}e 18, A-8700 Leoben, Austria}

\begin{abstract}
Quantum mechanical calculations had been previously applied to predict phase stability  in many ternary and multinary nitride systems.
While the predictions were very accurate for the Ti-Al-N system, some discrepancies between theory and experiment were obtained in the case of other systems.
Namely, in the case of Ta-Al-N, the calculations tend to overestimate the minimum Al content necessary to obtain a metastable solid solution with a cubic structure.
In this work, we present a comprehensive study of the impact of vacancies on the phase fields in quasi-binary TaN-AlN and NbN-AlN systems.
Our calculations clearly show that presence of point defects strongly enlarges the cubic phase field in the TaN-AlN system, while the effect is less pronounced in the NbN-AlN case.
The present phase stability predictions agree better with experimental observations of physical vapour deposited thin films reported in the literature than that based on perfect, non-defected structures.
This study shows that a representative structural model is crucial for a meaningful comparison with experimental data.
\end{abstract}


\end{frontmatter}

\section{Introduction}
\begin{figure}[ht]
  \centering
  \includegraphics[width=\singleColumnWidth]{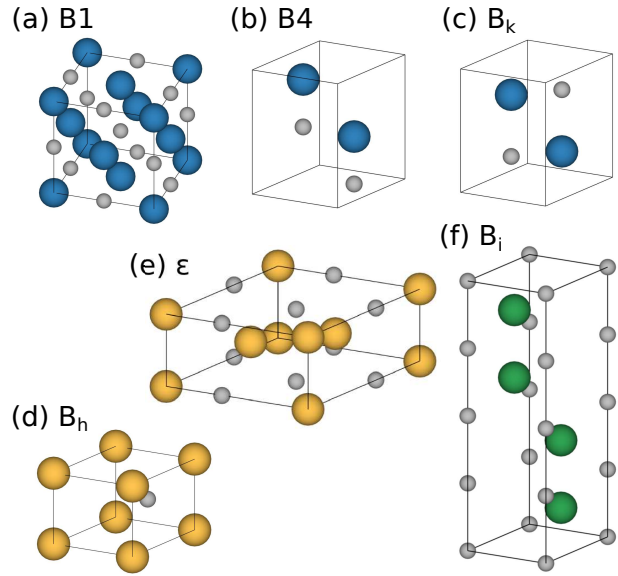} 
  \caption{Structures considered in the present work: (a) cubic rock-salt B1, (b) hexagonal wurtzite B4, (c) hexagonal \Bk, (d) hexagonal \Bh, (e) \eps, and (f) \Bi phases. Smaller grey and larger coloured spheres represent nitrogen and metal atoms, respectively. 
  The colour of the metal represent system for which a given phase was considered: orange for the Ta-Al-N system, green for the Nb-Al-N system, and blue for both systems. 
  Visualised using the VESTA software \cite{Momma2011-jk}.}
  \label{fig:structures}
\end{figure}

Phase stability of multinary nitride systems predicted using \textit{ab initio} methods has been repeatedly used for guiding experiments.
In the context of protective coatings the aim is usually to stabilise the cubic B1 phase (rock-salt, NaCl prototype, space group \#225, Fig.~\ref{fig:structures}a).
In particular, numerous studies on the quasi-binary TiN-AlN system showed the great degree of agreement between theory and experiment \cite{Mayrhofer2006-uc, Alling2007-bj, Zhang2007-gq, Alling2008-of, Hoglund2010-br, Rovere2010-qy, Holec2011-la}, but the same approach has been applied also to CrN--AlN \cite{Alling2007-tc, Alling2008-of, Mayrhofer2008-jb, Rovere2010-qy, Zhou2013-ze}, ZrN-AlN \cite{Sheng2008-yt, Rovere2010-qy, Holec2011-la}, HfN-AlN \cite{Alling2008-of, Rovere2010-qy, Holec2011-la}, ScN-AlN \cite{Hoglund2010-br, Hoglund2010-vm, Rovere2010-qy, Zhang2013-op}, NbN-AlN \cite{Holec2010-nb, Rovere2010-qy}, various TiN-TMN \cite{Zhang2008-wx, Abadias2013-nf} and CrN-TMN (TM=transition metal) systems \cite{Zhou2013-zx, Zhou2016-ev}, and multinary systems \cite{Hollerweger2016-mm, Riedl2013-ra, Lind2011-ol, Rachbauer2010-hc, Mayrhofer2010-ae, Flink2008-go, Rachbauer2012-tk, Rachbauer2011-lr, Rachbauer2012-uv}. 

The good agreement between the quantum-mechanical predictions and experiment is somewhat surprising considering the differences between simplified theoretical model and complex real materials.
Firstly, the calculations are done at $0\uu{K}$ unlike the finite temperature of the  reality.
\corr{Secondly, the bulk calculations are often compared with experimental data for thin films.
Thirdly}, defects are typically neglected in the idealised computational models while they are unavoidably present in the experimental samples, in particular in those synthesised using physical vapour deposition (PVD) techniques.

The finite temperature effects can be included at several levels. 
The simplest is to include configurational entropy.
As long as only relative stability at fixed composition of various polymorphs is concerned, however, the configurational entropy contribution is the same for all phases, and hence does not alter the phase stability.
The next important contribution is the entropy related to lattice vibrations.
Apart from semi-empirical formulae for estimating the Debye temperature, $\theta_D$, and a subsequent evaluation of the free energy within the Debye model \cite{Moruzzi1988-sd, Shang2010-qa}, also an explicit calculation using the finite-displacement method can be employed.
Despite the recent progress in available computational resources and efficiency of the methods, a direct calculation of the vibrational entropy remains computationally prohibitive for typical supercell sizes used to model solid solutions (50--100 atoms).
\corr{These sizes also imply, that the PVD thin films, being typically hundreds of nm up to several $\mu$m thick, can be realistically modelled with bulk materials from the {\itshape ab initio} perspective.}

Regarding imperfections of real materials, point defects such as vacancies, interstitials, anti-sites, Frenkel pairs, and Schottky defects, are computationally affordable.
The impact of vacancies on phase stability in the Ti-Al-N has been recently addressed by Euchner and Mayrhofer \cite{Euchner2015-ah} who concluded that while the presence of vacancies obviously alters energies of both competing cubic B1 and hexagonal B4 (wurtzite, ZnS prototype, space group \#186, Fig.~\ref{fig:structures}b) structures, it has only a negligible impact on the maximum Al solubility in the B1 phase, determined by the cross-over of the compositionally dependent energies of both phases.
We note that the vacancies have nevertheless a strong effect on the decomposition of the metastable cubic Ti$_{1-x}$Al$_x$N$_z$ phase \cite{Alling2008-ha}. 
\corr{This concept has been recently demonstrated experimentally by intentionally creating off-stoichiometric Ti$_{1-x}$Al$_x$N$_z$ thin films with vacancy concentrations over 10\% on either metal or N sublattice \cite{To_Baben2017-rd}.
Similar off-stoichiometry has been reported also for Ti-Al-N films synthesised using chemical vapour deposition \cite{Zalesak2017-lo}.}
In contrast to the Ti-Al-N system, vacancies and other point defects have been shown to change the phase stability in the Al-Cr-O \cite{Koller2016-ik} and Ti-W-B \cite{Euchner2015-kg} systems.

The ground state of TaN is a hexagonal \eps-phase (sometimes also denoted as $\pi$-phase, space group \#189, Fig.~\ref{fig:structures}e) \cite{Grumski2013-ax}.
A hexagonal \Bh-phase (sometimes also denoted as $\theta$-phase, space group \#187) is a metastable polymorph of TaN with energy of formation being $\approx36\uu{meV/at.}$ higher than that of \eps-TaN \cite{Koutna2016-ul}.
As mentioned before, the stable phase of AlN is a hexagonal wurtzite phase. 
Interestingly, the (metastable) B1 cubic phase is obtained using PVD for a broad range of compositions in the quasi-binary TaN-AlN systems \cite{Zhang2011-hi} despite both binary boundary systems being only metastable in this modification.
First principles calculations predict a minimum of $\approx0.2$ Al fraction on the metal sublattice of the Ta$_{1-x}$Al$_x$N system in order to stabilize the cubic B1 phase \cite{Holec2013-wm}.
Experimentally, however, the cubic phase has been obtained for much lower Al contents using PVD \cite{Zhang2011-hi}.
Since TaN has been shown to prefer vacancies in its cubic modification \cite{Stampfl2003-yh, Koutna2016-kk}, the question arises whether this discrepancy between first principles calculations and experimental data might be caused by neglecting the point defect in the simulations.
Additionally, an isovalent system, NbN-AlN, will be investigated here as it also exhibits similar behaviour in terms of the boundary binary systems being stable in hexagonal phases (ground state of NbN is the \Bi-phase, space group \#194, Fig.~\ref{fig:structures}f), while the cubic phase is obtained for Nb$_{1-x}$Al$_x$N \cite{Holec2010-nb}.

\section{Methods}
\subsection{First principles calculations}
The calculations were carried out using Vienna Ab initio Simulation Package (VASP) \cite{Kresse1996-gt}, a plane-wave implementation of the density functional theory \cite{Hohenberg1964-in, Kohn1965-rd}.
The electron-ion interactions were modelled using projector-augmented-method capable pseudopotentials \cite{Kresse1999-if}, while the electron-electron quantum-mechanical exchange and correlation interactions were described using the generalised gradient approximation as parametrised by Wang and Perdew \cite{Wang1991-ca}.
The plane-wave cut-off energy of $450\uu{eV}$ and $k$-point separation of $0.017\uu{\AA}^{-1}$ guarantee total energy accuracy of a few meV/at.
A structural relaxation involving optimisation of the supercell volume, shape, and atomic positions was stopped once the total energy fluctuations were smaller than $\approx0.005\uu{meV/at.}$

The considered structures were described in the Introduction and are visualised in Fig.~\ref{fig:structures}. 
Additionally, also a \Bk structure (space group 194, Fig.~\ref{fig:structures}c) was included in the analysis, as it represents a symmetry-stabilised transition state between wurtzite B4 and cubic B1 phases.
The chemical disorder of solid solutions was modelled using a supercell approach employing the special quasi-random structures (SQSs) \cite{Wei1990-zt}.
System sizes were chosen so that the resulting supercells contained $\approx60$--$70$ atoms.
In particular, $2\times2\times2$ supercells with 64 atoms based on the cubic B1 conventional cell, $3\times3\times2$ supercells with 72 atoms based on the B4 unit cell, $3\times3\times2$ supercells with 72 atoms based on the \Bk unit cell, $3\times3\times3$ supercells with 54 atoms based on the \Bh unit cell, $2\times2\times3$ supercells with 72 atoms based on the \eps-phase unit cell, and $3\times3\times1$ supercells with 72 atoms based on the \Bi unit cell, were employed.
When constructing the SQSs, pair interactions up to the seventh coordination shell were optimised to best match those of a random solid solution with the same composition.
Vacancies were treated as a ternary alloying element during the SQS generation.

\subsection{Free energy surfaces}
Energy of formation, $E_f$, is used to compare the (chemical) stability of various phases. 
It is defined as
\begin{equation}
  E_f=\frac{E_{\mathrm{tot}}(\xi) - \sum_i N_i\mu_i}{\sum_i N_i}\ ,
\end{equation}
where $E_{\mathrm{tot}}(\xi)$ is the total energy (per supercell) of a compound $\xi$ with a composition given by numbers $N_i$ of species $i$ (Ta/Nb, Al, N). 
$\mu_i$ is chemical potential (per atom) of a species $i$ set to bcc-Ta, bcc-Nb, fcc-Al, and N\textsubscript{2} molecule. 
The resulting energy of formation expresses energy gain (when negative) per atom by creating a crystal (here solid solution) out of individual elements in a thermodynamic equilibrium with respective crystalline (Ta, Nb, Al) and molecular (N) reservoirs.
It should be noted that the actual choice of reference state does not influence relative stability of phases at fixed composition.

In order to unambiguously represent constitution (i.e., chemical composition and vacancy content) of a compound $\xi$, $\textrm{(TM}_{1-x}\textrm{Al}_{x}\textrm{)}_{1-y}\textrm{N}$ and $\textrm{TM}_{1-x}\textrm{Al}_{x}\textrm{N}_{1-z}$ notations will be used for metal- and nitrogen-vacancy cases, respectively.
Here, TM is either Ta or Nb, $x$ denotes the aluminium fraction of the metal species (Al/(TM+Al)), and $y$ or $z$ indicate the vacancy content on the metal or on the nitrogen sublattice, respectively.

The $E_f(\xi)$ values were calculated for the sets of discrete atomic compositions related to the unit cell geometry and the supercell size, both being in general different for different phases, which in turn makes a directed comparison of energies of various phases impossible.
To create a phase mapping of the most stable structures, continuous functions $E_f(x,y)$ and $E_f(x,z)$ were constructed for each phase by fitting the discrete data points from the \textit{ab initio} calculations using a below described multi-step method. 

$E_f(x,y|z)$ assumes following functional form
\begin{equation}
  E_f(x,y|z) = a_0(y|z) + x\,a_1(y|z) + x^2\,a_2(y|z) + x^3\,a_3(y|z)\ , 
  \label{totalFit}
\end{equation}
i.e., it is a third order polynomial in the aluminium fraction, $x$, with coefficients $a_i(y|z)$ depending on the metal ($y$) or nitrogen ($z$) vacancy content.
In the first step, the \textit{ab initio} calculated data is used to fit coefficients $b_i(y|z)$ of several third order polynomials being functions of the aluminium concentration, each for a fixed vacancy content:
\begin{equation}
  p(x)\big|_{y|z=\mathrm{const.}} = b_0(y|z) + x\,b_1(y|z)x + x^2\,b_2(y|z) + x^3\,b_3(y|z)\ . 
  \label{p(x)}
\end{equation}
In the second step, the coefficients $b_i(y|z)$ were used to determine functions 
\begin{equation}
  a_i(y|z) = a^0_i + (y|z) a^{1,y|z}_i + (y|z)^2 a^{2,y|z}_i
  \label{a(y)}
\end{equation}
for every term $i$ in Eq. \ref{totalFit}. 
To maintain continuity of the fitted $E_f$ for $y,z\to0$, the parameters $a^0_i$ were set to be identical to the coefficients $b_i(y=z=0)$.
The other parameters, $a^{1,y|z}_i$ and $a^{2,y|z}_i$ are determined separately for $y$ and $z$ by minimizing the sum of the squared residuals of the function in Eq.~\ref{a(y)}.

This particular choice of the fitting procedure was motivated by maintaining the continuity between $E_{f}(x,y)$ and $E_{f}(x,z)$, providing reasonable flexibility to fit the \textit{ab initio} data (e.g., quadratic fit in $x$ or vacancy-content independent coefficients $a_i(y,z)$ did not provide representative fits). 
Splitting $E_f$ into two linked functions for $y$ and $z$ ensures that data with no causal contribution are not used for fitting, i.e., configurations with metal (nitrogen) vacancies do not provide any information about the impact of nitrogen (metal) vacancies on the structure. 

Once the functions $E_{f}(x,y)$ and $E_{f}(x,z)$ were found for every phase, the assessment of phase stability was carried out by identifying which structure yields the lowest $E_f$ value for each $(x,y|z)$.

\section{Results}
\subsection{Ta-Al-N system}
\begin{figure}[ht]
  \centering
  \parbox{\singleColumnWidth}{
  (a) metal vacancies\newline
  \includegraphics[width=\singleColumnWidth]{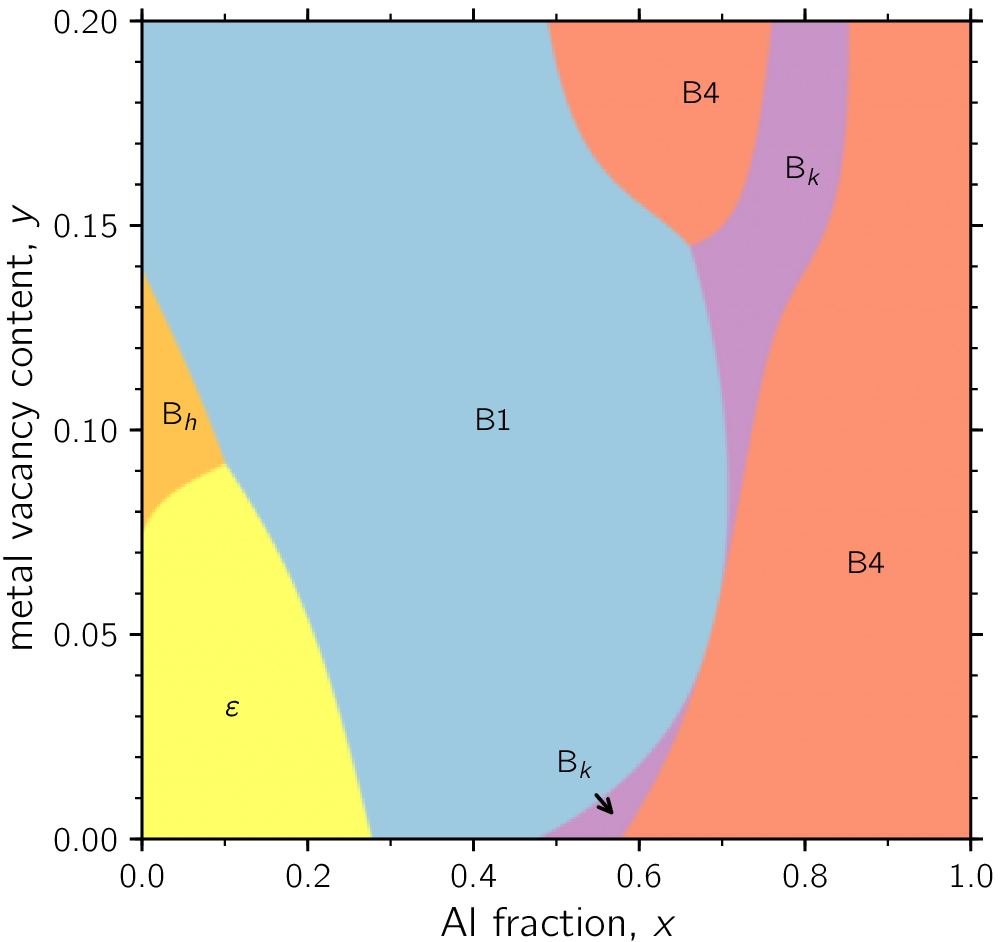}
  }  
  \parbox{\singleColumnWidth}{
  (b) nitrogen vacancies\newline
  \includegraphics[width=\singleColumnWidth]{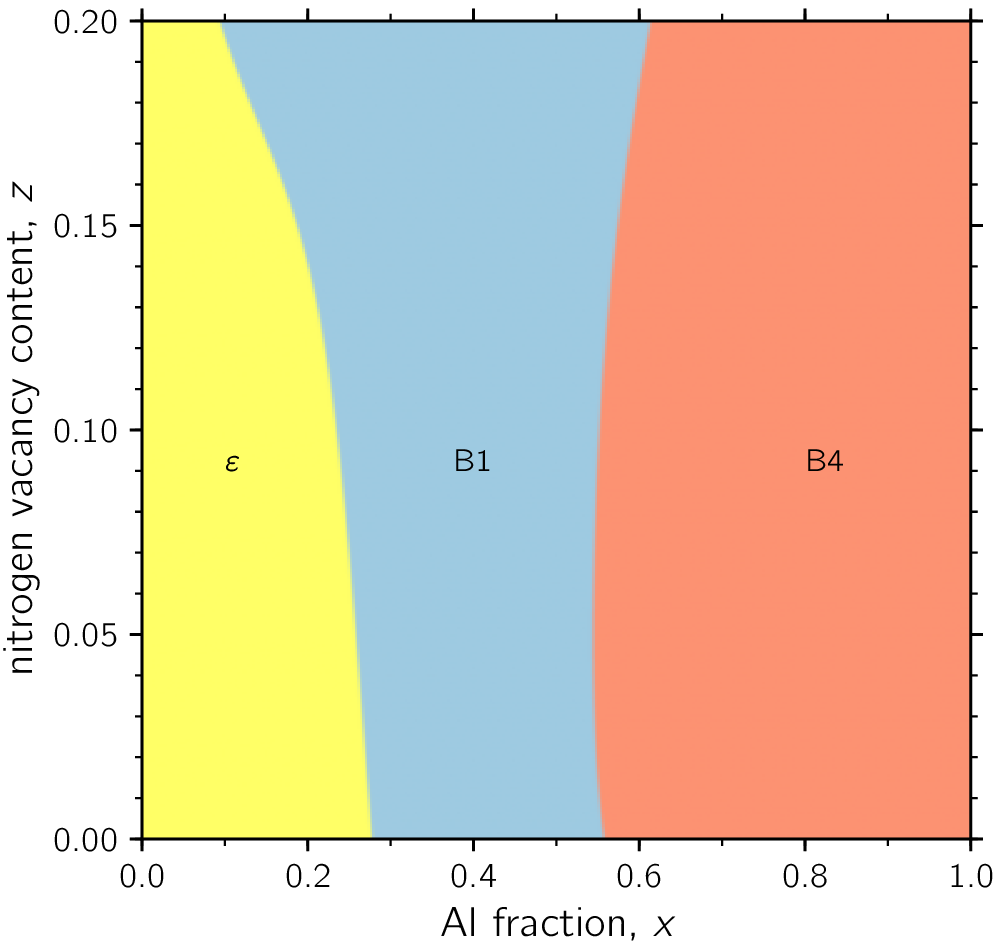}
  }
  \caption{Estimated single-phase fields of the most stable solid solution for the Ta-Al-N system as a function of the Al/(Al+Ta) fraction, $x$, and the content of vacancies on (a) the metal and (b) the nitrogen sublattice.}
  \label{fig:phaseDiag-TaAlN}
\end{figure}

Considering only the perfect structures ($y=0$ in Fig.~\ref{fig:phaseDiag-TaAlN}a or $z=0$ in Fig.~\ref{fig:phaseDiag-TaAlN}b), for Al fractions up to $\approx0.27$ the hexagonal \eps-phase is predicted as the most stable one.
For higher Al contents $0.27\lessapprox x\lessapprox0.48$, the cubic B1 phase is the most stable one, followed by the hexagonal \Bk phase for $0.48\lessapprox x\lessapprox0.58$, and finally the Ta-Al-N system is predicted to prefer the wurtzite B4 phase for $x\gtrapprox0.58$.

With the increasing amount of metal vacancies, the maximum solubility of Al in the \eps-phase decreases; the \eps-phase is substituted by the \Bh phase for $y\gtrapprox0.09$, and the cubic B1 phase become favourable even for pure TaN at $y\approx0.14$ ($14\%$ of the metal sites being vacant). 
Similarly, the hexagonal \Bk/B4 phase fields shrink for $y$ increasing up to $\approx0.1$ (at which the maximum Al solubility, $x\approx0.7$, in the cubic phase is reached). 
For an even higher content of the metal vacancies, the cubic-to-hexagonal transition shift back to lower $x$.

The cubic phase field expands also with the increasing amount of the vacancies on the nitrogen sublattice (Fig.~\ref{fig:phaseDiag-TaAlN}b), although the impact is not so strong as in the case of metal vacancies. 
We note that the \Bk phase was not explicitly considered in this case, however, the B4 phase could structurally relaxed into the \Bk phase if this was energetically preferred (the \Bk phase is higher-symmetry variant of the B4 phase, cf. Figs.~\ref{fig:structures}b and c).

\begin{figure}[ht]
  \centering
  \includegraphics[width=\singleColumnWidth]{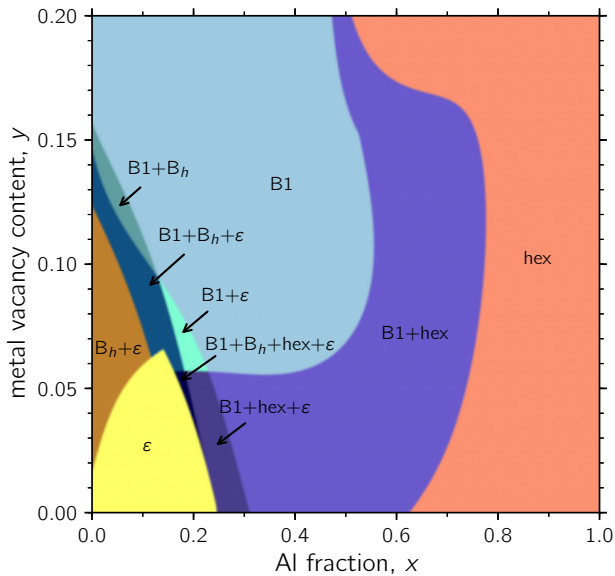}
  \caption{Multi-phase fields of the most stable and metastable up to $\Delta E_f\leq+25\uu{meV/at.}$ solid solutions for the Ta-Al-N system, as a function of the Al/(Al+Ta) fraction, $x$, and the content of vacancies on the metal sublattice.}
  \label{fig:phaseDiag-TaAlN-extended}
\end{figure}

An inspection of the actual $E_f$ surfaces revealed that, in several regions, more phases were energetically close and hence could co-exist at finite temperatures, similarly to the phase stability considerations in Zr-Al-N and Hf-Al-N systems \cite{Holec2011-la}.
Figure \ref{fig:phaseDiag-TaAlN-extended} shows an extension of the single-phase diagram in Fig.~\ref{fig:phaseDiag-TaAlN}a for two- and more-phase regions.
The co-existence is limited to energy difference $\Delta E_f\leq25\uu{meV/at.}$ from the lowest-energy structure ($25\uu{meV/at.}$ is approximately the value of the Boltzmann factor $k_BT$ at the room temperature).
The single-phase cubic B1 field disappears completely for metal vacancy content below $y\approx0.05$, where it is predicted to co-exist mostly with the hexagonal \Bk or B4 phases which were merged for this analysis due to their structural similarity.
For a low Al content and $y\lessapprox0.07$, the single-phase field of the \eps-phase exists; for higher metal vacancy content, a competition of the hexagonal \eps and \Bh phases is predicted.
The hexagonal \Bh/B4 single-phase field exists for $x\gtrapprox0.65$ for perfect structures, and the field shrinks with increasing $y$.
Importantly, for $y\gtrapprox0.05$, a wide single phase \corr{B1} field between \corr{$\approx0.2$} and $\approx0.55$ opens, which further widens for increasing metal vacancy content.

\subsection{Nb-Al-N system}
\begin{figure}[ht]
  \centering
  \parbox{\singleColumnWidth}{
  (a) metal vacancies\newline
  \includegraphics[width=\singleColumnWidth]{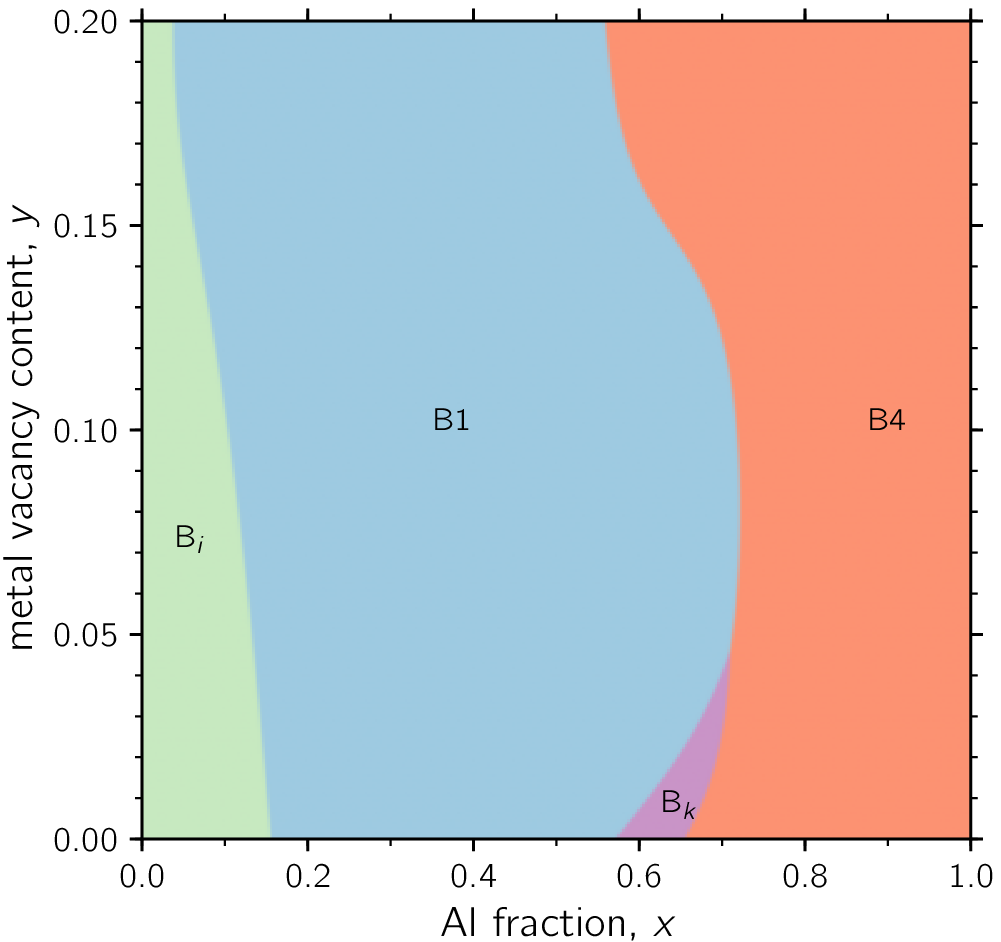}
  }
  \parbox{\singleColumnWidth}{
  (b) nitrogen vacancies\newline
  \includegraphics[width=\singleColumnWidth]{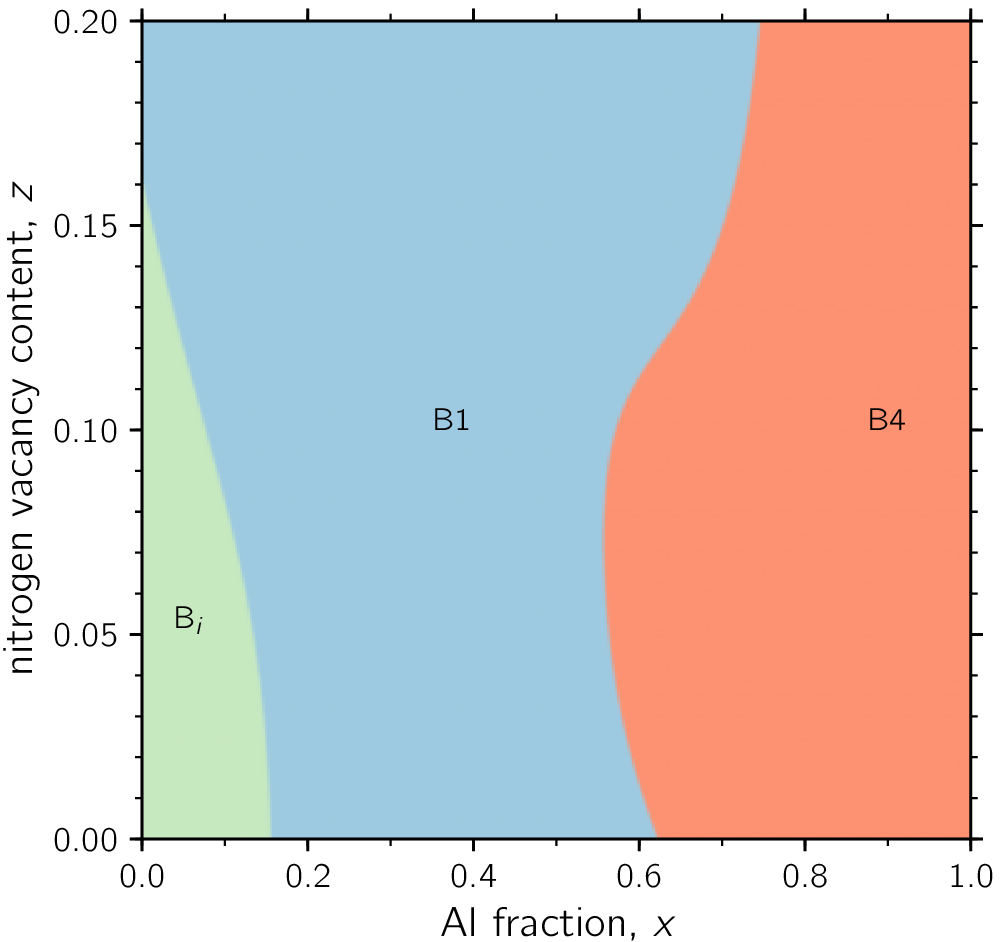}
  }
  \caption{Estimated single-phase fields of the most stable solid solution for the Nb-Al-N system, as a function of the Al/(Al+Nb) fraction, $x$, and the content of vacancies on (a) the metal and (b) the nitrogen sublattice.}
  \label{fig:phaseDiag-NbAlN}
\end{figure}

The evaluated single-phase field diagrams for the Nb-Al-N system with vacancies on the metal and nitrogen sublattices are shown in Figs.~\ref{fig:phaseDiag-NbAlN}a and b, respectively.
The \Bi phase yields the lowest $E_f$ for Al fractions $x\lessapprox0.16$ in the case of perfect NbN-AlN quasi-binary system, in agreement with previous calculations \cite{Holec2010-nb}.
Increasing metal ($y$) or nitrogen ($z$) vacancy content lowers the maximum Al content soluble in the hexagonal \Bi phase, and consequently extends the cubic B1 phase field.
The hexagonal \Bk/B4 phases are stable for $x\gtrapprox0.6$ for the perfect structures.
Their phase field shrinks (B1 phase field expands) with increasing metal vacancy content.
On the contrary, increasing $z$ leads to a slight expansion of the hexagonal phase field, and thus a slight shrinkage of B1 phase, before the situation reverses for $z\gtrapprox0.12$.

\begin{figure}[ht]
  \centering
  \includegraphics[width=\singleColumnWidth]{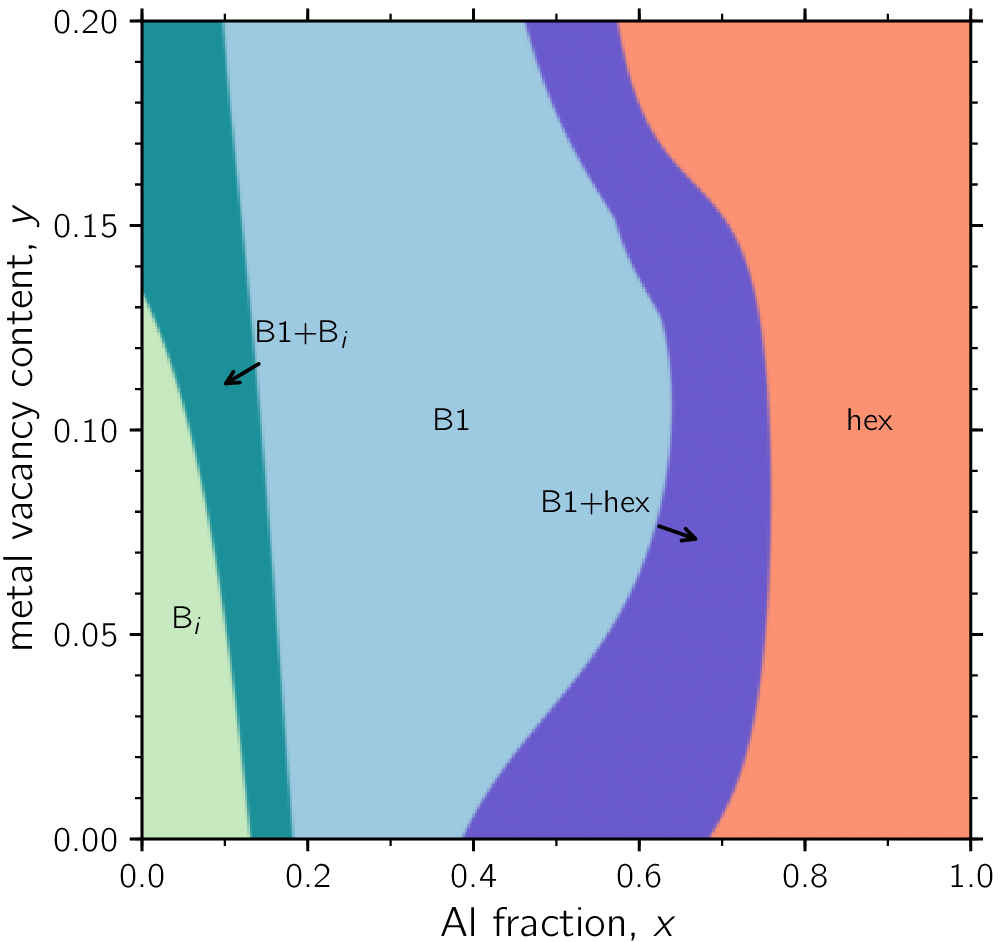}
  \caption{Multi-phase fields of the most stable and metastable up to $\Delta E_f\leq+25\uu{meV/at.}$ solid solutions for the Nb-Al-N system, as a function of the Al/(Al+Nb) fraction, $x$, and the content of vacancies on the metal sublattice.}
  \label{fig:phaseDiag-NbAlN-extended}
\end{figure}

Similarly to the Ta-Al-N, the multi-phase diagram was evaluated for $\Delta E_f\leq25\uu{meV/at.}$ also for the Nb-Al-N system (Fig.~\ref{fig:phaseDiag-NbAlN-extended}).
It follows that the single-phase \Bi field is reduced, and a non-negligible two-phase \Bi+B1 field appears.
In contrast to the Ta-Al-N system, a single phase cubic B1 field is predicted for $0.18\lessapprox x\lessapprox0.43$ even for the perfect configuration without vacancies, and it significantly expands with increasing metal vacancy content towards larger maximum Al solubility in the B1 phase.

\section{Discussion}
\subsection{Origin of the extended stability range of the cubic phase}
\begin{figure*}[p]
  \centering
  \includegraphics[width=0.9\textwidth]{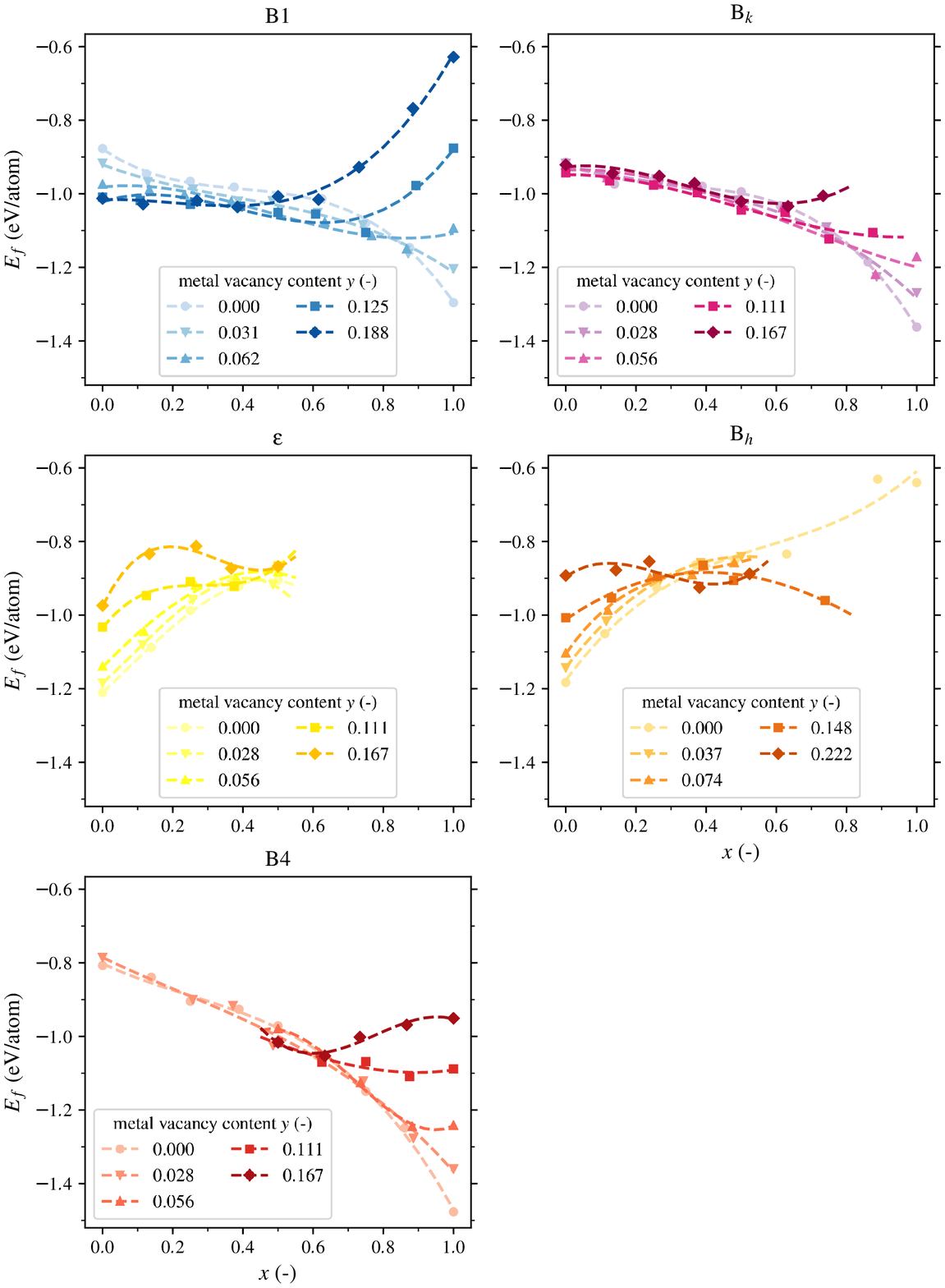}
  \caption{Energy of formation for various phases in the Ta-Al-N as a function of the Al/(Al+Ta) ratio and amount of vacancies on the metal sublattice. 
  Full symbols represent the actual DFT-calculated values while the dashed lines are the fitted polynomials according to Eq.~\ref{a(y)}.}
  \label{fig:Ef_TaAlN}
\end{figure*}

The large B1 single-phase fields in Figs.~\ref{fig:phaseDiag-TaAlN} and \ref{fig:phaseDiag-NbAlN}, as well as the various B1 containing multi-phase fields in Figs.~\ref{fig:phaseDiag-TaAlN-extended} and \ref{fig:phaseDiag-NbAlN-extended}, clearly show that the stability region of the cubic phase generally expands by increasing the vacancy content.
To shed some light on the origin of this stabilization, the raw $E_f$ data are plotted in Fig.~\ref{fig:Ef_TaAlN} for each individual phase in the Ta-Al-N system with metal vacancies.

While the \Bh and \eps phases exhibit a considerable increase in $E_f$ for increasing $y$ values at low aluminium fractions $x$ (i.e., metal vacancies are undesirable structural perturbation, a defect in the very sense of the term), the formation energy of B1 becomes more negative with increasing $y$, making the structure more stable. 
This is in agreement with calculations by other authors of pure TaN, which showed that the cubic phase favours vacancies \cite{Stampfl2005-xt, Grumski2013-ax, Koutna2016-kk}.
The decrease in $E_f$ extends up to $x\approx0.5$ where the metal vacancies eventually have an adverse effect. 
Contrarily, the \Bk phase is affected by the metal vacancies only a little, in particular for small vacancy content $y\lessapprox0.07$.
For low aluminium fractions $x$, $E_f$ monotonously increases with increasing $y$, but surprisingly the metal vacancies seem to stabilise the phase for $0.4\lessapprox x\lessapprox0.55$.
The energy of formation of B4 increases with increasing $y$ in the compositional range of interest ($x\gtrapprox0.6$). 

Interestingly, the curvature of the $E_f(x)$ functions of the B1 phase changes gradually from concave to convex with increasing $y$-values (Fig. \ref{fig:Ef_TaAlN}). 
This means that the defected solid solution becomes chemically stable for aluminium fractions $0<x<1$, and loses its internal driving force for an isostructural decomposition.

The structural relaxations of some of the B4, \Bh and \eps cases with compositions around $x\approx0.5$ were extremely difficult to converge, in particular for higher vacancy concentrations. 
This, in addition to the high $E_f$ values, indicates, that the respective configurations experience significant forces acting on atoms and the simulation cell, hence leading to significant distortions away from the nominal hexagonal phases. 

Although the B1 phase field is slightly expanded also by an increased vacancy content $z$ on the nitrogen sublattice, the effect is much less pronounced than in the case of metal vacancies. 
This is because N vacancies are favourable in the B1 phase only for very low Al fractions $x\lessapprox0.1$, in agreement with recent studies for pure TaN \cite{Koutna2016-kk}.
For all other compositions and phases, $E_f$ increases.
The increase is largest for the \Bk phase
and for the wurtzite B4 phase (hence the B1 phase field expands to larger $x$ with increasing $z$).

Finally we note that the behaviour of the Nb-Al-N system is qualitatively the same, although the stabilisation of the cubic B1 phase by metal vacancies is not as pronounced as in the Ta-Al-N system.
The lowest $E_f$ is obtained for $y\approx0.06$ and it increases again for higher amount of metal vacancies.

\subsection{Experimental validation}

\begin{figure}
    \centering
    \includegraphics[width=\singleColumnWidth]{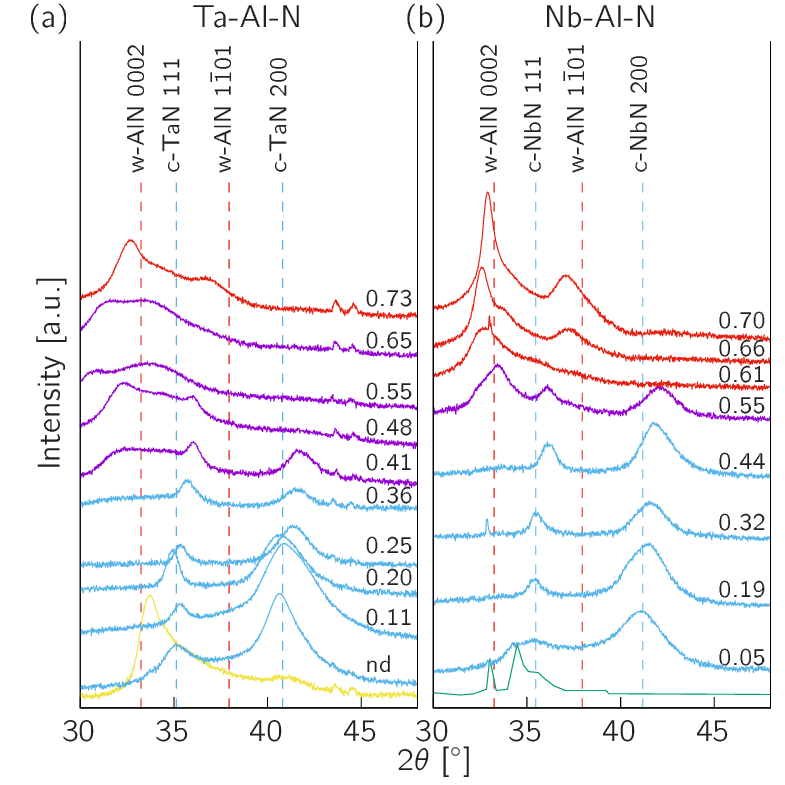}
    \caption{Experimental x-ray diffractograms of (a) Ta-Al-N and (b) Nb-Al-N systems as functions of Al metal-fraction $\mathrm{Al}/(\mathrm{Al}+M)$, $M=$Ta, Nb (``nd'' stands for non-detectable).
    The red, purple and blue curves denote single phase hexagonal wurtzite-like structure, dual phase hexagonal and cubic structure, and single-phase cubic structure, respectively.
    Binary TaN and NbN crystallised in their respective \eps and \Bi phases, respectively.
    Data reproduced from Ref.~\cite{Zhang2011-hi}.}
    \label{fig:XRD}
\end{figure}

Besides the hexagonal \eps-phase, the cubic B1 structure has been frequently found in pure TaN synthesised by various methods \cite{Taniguchi2014-xq, Marihart2015-fm, Zhang2011-hi, Shin1999-lp, Tsai1996-er, Kim2004-wq}. 
The cubic B1 phase in the Ta-Al-N system was experimentally found as a single-phase up to aluminium fractions $x=0.36$, and as a two-phase mixture with B4 up to $x=0.65$ \cite{Zhang2011-hi, Koller2016-rf, Chen2016-wj} (see Fig.~\ref{fig:XRD}a), which agrees well with the here predicted stability regions.
The Ta-Al-N thin films investigated by Koller \textit{et al.} \cite{Koller2016-rf} had an overall chemical formula of Ta$_{0.89}$Al$_{0.11}$N$_{1.2}$. 
Attributing the nitrogen over-stoichiometry solely to vacancies on the metal sub-lattice yields $y=0.167$ in the notation used here. 
This composition lies well within the B1 phase field according to the phase diagram in Fig.~\ref{fig:phaseDiag-TaAlN-extended}. 

The cubic B1 structure has been reported by various authors also in the Nb-Al-N system. 
Several works dealt with the structure of near-to-stoichiometric NbN films and reported on co-existence and phase transformations of hexagonal \Bi and the cubic B1 phase \cite{Zhitomirsky1998-cc, Benkahoul2004-pe, Oya2008-yn}. 
Regarding the quasi-binary system, Selinder \textit{et al.} reported a single-phase cubic structure in reactively sputtered thin films up to aluminium fractions $x\approx0.5$, and a two-phase structure together with the wurtzite B4 phase up to $x\approx0.6$ \cite{Selinder1995-yp}.
The same trend was obtained also by Zhang \cite{Zhang2011-hi} (see Fig.~\ref{fig:XRD}b).
Experiments by Franz \textit{et al.} yielded similar results for cathodic arc evaporated materials with mainly cubic structure up to $x\lessapprox0.56$ depending on the bias voltage \cite{Franz2010-ax}. 
Their chemical analysis revealed a slight nitrogen over-stoichiometry, which, when interpreted as being caused by a few at.\,\% of metal vacancies, leads to the enhancement cubic B1 phase field (Fig.~\ref{fig:phaseDiag-NbAlN}a).
Nevertheless, the multiphase diagram in Fig.~\ref{fig:phaseDiag-NbAlN-extended} suggest a strong phase competition with the hexagonal \Bk and B4 phases at the Al-rich end of the B1 phase field, $x\approx0.55$.

Although these qualitative, and often quantitative, agreements of our predictions with the available data are very encouraging, it is important to realise that the point defects, despite being an important element in the phase stability assessment, are only one of several simultaneously acting effects.
For example, increasing bias voltage during synthesis enhances the cubic phase field as demonstrated by Franz \textit{et al.} \cite{Franz2010-ax}.
This effect can be explained by increased amount of compressive stresses caused by high-energy of the impinging particles, which leads to stabilisation of the cubic phase due to its smaller specific volume \cite{Alling2009-ou, Holec2010-mo}.
Similarly, we anticipate that a full account for the phase stability predictions of PVD synthesised materials must include also impact of other defects (e.g., interstitials, anti-sides, etc.) as well as limited diffusion kinetics related to growth rate.

\section{Conclusions}
Extensive first principles calculations of phase stability in defected quasi-binary TaN-AlN and NbN-AlN systems were performed.
A special focus was laid on the impact of vacancies, representatives of point defects unavoidably present in materials due to thermodynamic reasons as well as due to deposition process itself.
Our calculations clearly show that the phase field of the rock-salt cubic phase expands with increasing amount vacancies in both material systems.
The B1 phase field extends towards lower as well as towards higher Al content on the metallic sublattice for Ta-Al-N, whereby the impact of the metal vacancies is stronger than that of the N vacancies for vacancy content up to $\approx15\%$ on the respective sublattice.
The same applies also to metal vacancies in Nb-Al-N, while the width of the cubic phase field is not significantly affected by the N vacancies for sublattice vacancy content up $\approx10\%$.
The underlying driving force for the cubic phase stabilisation was tracked to the natural tendency of c-TaN and c-NbN to contain vacancies (unlike the other phases) and to the smallest specific volume of the B1 phase. 
In summary, by including the impact of point defects in the phase stability assessment, we obtained predictions which are significantly closer to experimental results on metastable PVD-synthesised thin films.

\section*{Acknowledgements}
The computational results presented have been achieved using the Vienna Scientific Cluster (VSC).
The financial support by the Austrian Federal Ministry of Economy, Family and Youth and the National Foundation for Research, Technology and Development is gratefully acknowledged.

\section*{References}
\bibliographystyle{elsarticle-num} 
\bibliography{refs}

\end{document}